\documentclass[a4paper,12pt]{article}

\usepackage{graphics}
\newcommand{\ket}[1]{\ensuremath{\left| #1 \right\rangle}}

\newcommand{\br}[1]{\ensuremath{\left\langle #1 \right.}}

\newcommand{\bk}[2]{\br{{#1}}\ket{{#2}}}

\newcommand{\realpict}[2]
{
\begin{figure}[htb]
\includegraphics{#1}
\caption{#2\label{fg:#1}}
\end{figure}
}

\begin{document}
\title{Are all reversible computations tidy?}
\author{O J E Maroney\\
HH Wills Physics Laboratory\\
University of Bristol\\
Tyndall Avenue \\
Bristol \\
BS8 1TL\\
o.maroney@bristol.ac.uk}
\date{}

\maketitle
\begin{abstract}
\label{abstract}

It has long been known that to minimise the heat emitted by a 
deterministic computer during it's operation it is necessary to 
make the computation act in a logically reversible 
manner\cite{Lan61}. Such logically reversible operations require a 
number of auxiliary bits to be stored, maintaining a history of 
the computation, and which allows the initial state to be 
reconstructed by running the computation in reverse. These 
auxiliary bits are wasteful of resources and may require a 
dissipation of energy for them to be reused. A simple procedure 
due to Bennett\cite{Ben73} allows these auxiliary bits to be 
"tidied", without dissipating energy, on a classical computer. All 
reversible classical computations can be made tidy in this way. 
However, this procedure depends upon a classical operation 
("cloning") that cannot be generalised to quantum 
computers\cite{WZ82}. Quantum computations must be logically 
reversible, and therefore produce auxiliary qbits during their 
operation. We show that there are classes of quantum computation 
for which Bennett's procedure cannot be implemented. For some of 
these computations there may exist another method for which the 
computation may be "tidied". However, we also show there are 
quantum computations for which there is no possible method for 
tidying the auxiliary qbits. Not all reversible quantum 
computations can be made "tidy". This represents a fundamental 
additional energy burden to quantum computations.  This paper 
extends results in \cite{Mar01}.

\end{abstract}

\section{Reversible Classical Computation}\label{classical}

A reversible calculation may be defined as one which operates, 
upon an input state $i$ and an auxiliary system, prepared in an 
initial 'known' state $Aux0$ , to produce an output from the 
calculation $O(i)$, and some additional information $Aux(i)$:

\[
F:(i,Aux0)\rightarrow (O(i),Aux(i))
\]

in such a manner that there exists a complementary calculation:

\[
F^{\prime }:(O(i),Aux(i))\rightarrow (i,Aux0)
\]

Given that the mapping $i \rightarrow O(i)$ will, for 
deterministic computation, be many-to-one, the existence of the 
auxiliary output bits is necessary to uniquely specify the input 
state from the output state, so that the computation can be 
reversed.  They may be considered to be a 'history' of the 
computation that retains the information that an irreversible 
computation would have discarded.

A general procedure for discovering the complementary calculation 
$F^{\prime }$ can be given like this:
\begin{itemize}
\item Take all the logical operations performed in $F$, and reverse
their operation and order.
\end{itemize}
As long as all the logical operations in $F$ are reversible logic 
gates, this is possible. It is known that the reversible 
Fredkin-Toffoli gates are capable of performing all classical 
logical operations, so it is always possible to make a computation 
logically reversible. 

However, they also represent wasted resources which we would like 
to recover, to be used in further calculations.  One way to do 
this would be to reset the bits using Landauer 
Erasure\cite{Lan61}.  However, this would require work to be 
performed upon the system which would be dissipated as heat.  
Another way to recover the bits would be to reverse the 
computation, using $F^{\prime}$, but this would lose us all 
record of the output of the computation, so is of little use.

Bennett\cite{Ben73} showed that a solution was to find a 
different reverse calculation $F^{\prime\prime}$

\[
F^{\prime \prime }:(O(i),Aux(i),AuxO)\rightarrow (i,Aux0,O(i))
\]

The reconstruction of the original input states $i$ is sufficient 
to secure reversibility.  A general procedure for 
$F^{\prime\prime}$, is:
\begin{itemize}
\item Copy $O(i)$ into a further auxiliary system $AuxO$ by means of
a Controlled-NOT gate;
\item Run $F^{\prime }$ on the original system.
\end{itemize}
This has also been shown to be the optimal 
procedure\cite{LTV98,LV96} for $F^{\prime\prime}$. We call such a 
calculation TIDY. All classical reversible computations can be 
made TIDY.

The great strength of Bennett's procedure is that it does not 
require any calculation of what are the specific, determinate but 
'unknown', states of the auxiliary bits or the output states.  
Although they will be some determinate function of the input 
states, (eg. of the form {\em (A XOR B) AND (C OR D)} ) we cannot 
use this {\it directly }to reset the auxiliary states 
non-dissipatively (eg. by calculating whether the state is zero 
or one and performing a NOT operation upon it if it is one).  To 
construct such an augmented computation, we would first have to 
calculate what the outputs would be for each input state and this 
is just to run the calculation $F$ itself!
\section{Reversible Quantum Computation}\label{quantum}
When we attempt to apply Bennett's procedure to quantum 
computations, we find it can fail.  The cause is well known. The 
problem is that the Controlled-NOT gate does not act as a 
universal copying gate for quantum computers. In fact, the 
universal copying gate does not exist, as a result of the 
'no-cloning theorem'\cite{WZ82}.

Clearly, in the case where the output states from a quantum 
computer are in a known orthogonal set, then the quantum 
computation can be made tidy. This is suggestive not of a general 
quantum computation, but of limited quantum algorithmic boxes: 
each connected by classical communication. Distributed quantum 
information may be desirable - in particular, a more general 
conception of quantum computation may be required which takes 
inputs from different sources, and/or at different 
times.\realpict{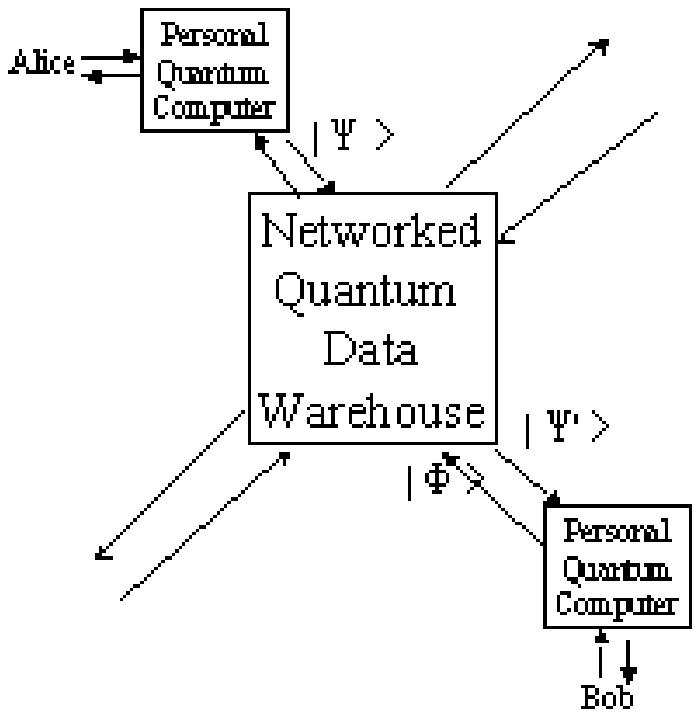}{Distributed quantum computing} In 
Figure \ref{fg:qnetwork} we see an example of this - Alice 
performs some quantum computation, and stores the result of it in 
a 'quantum data warehouse'. At some later time, Bob takes part of 
these results as an input into his own computation.

We are going to take our definition of a quantum 
computation\footnote{There is further complication when 
entanglement enters the problem. When the output part of an 
entangled state is non-recoverably transmitted, the loss of free 
energy in the remainder is always at least equal to $kT$ times 
the entropy of the reduced density matrix of the output. However, 
this minimum loss of free energy requires knowledge of an accurate 
representation of the resulting density matrix - which may not be 
possible without explicitly calculating the output states.} as the 
operation:

\[
U_C:\ket{i}\ket{Aux0}\rightarrow \ket{O(i)}\ket{Aux(i)}
\]

so that the output is always in a separable state.

In other words, we regard the 'output' of the computation as the 
subsection of the Hilbert space that is interesting, and the 
'auxiliary' as everything that is uninteresting. If the 'output' 
were entangled with the 'auxiliary' space, then there would be 
additional information relevant to the 'output', contained in the 
super-correlations between 'output' and 'auxiliary' spaces. In 
any event, such entanglement can only increase the energy lost.

As any quantum computation must be performed by a unitary 
operation, all quantum computers must be reversible. But are they 
TIDY?

If this model of computation is classical, then each time data is 
sent to the central database, the local user can copy the data 
before sending it, and tidy up their computer as they go along. 
At end of all processing - if it happens - reconstruction of 
computation from stored input would allow tidying of any stored 
data no longer needed. The difference between computation using 
distributed classical algorithmic boxes and a single classical 
computation is a trivial distinction, as the computation may be 
tidied up along the way. However, this distinction depends upon 
the classical nature of the information transferred between the 
algorithmic boxes.

In our generalised quantum computation network, we can no longer 
guarantee that the operations performed at separate locations are 
connected by classical signals only. We now need to generalise the 
definition of tidiness to quantum computers.

Considering a general operation, unitarity requires that the inner 
products between different input states and between the 
corresponding output states is unchanged by the computation. 
Reversibility must always hold. This leads to the conditions:
\paragraph{Reversible}
\begin{eqnarray*}
\bk{i}{j}\bk{Aux0}{Aux0}&=&\bk{O(i)}{O(j)}\bk{Aux(i)}{Aux(j)}
\end{eqnarray*}
\paragraph{Tidy}
\begin{eqnarray*}
\bk{i}{j}\bk{Aux0}{Aux0}\bk{AuxO}{AuxO}
    &=&\bk{i}{j}\bk{O(i)}{O(j)}\bk{Aux0}{Aux0}
\end{eqnarray*}

We can eliminate $\bk{Aux0}{Aux0}=1$ and $\bk{AuxO}{AuxO}=1$, 
leaving only three cases.

\subsection{Orthogonal Outputs}

The output states are orthogonal set:

\[
\bk{O(i)}{O(j)}=\delta _{ij}
\]

Reversibility {\it requires} the input states to be an orthogonal 
set $\bk{i}{j}=0$, and the TIDY condition will hold. This is not 
too surprising, as an orthogonal set of outputs {\it can} be 
cloned, and so can be tidied using Bennett's procedure.

\subsection{Orthogonal Inputs}

The input states are an orthogonal set $\bk{i}{j}=\delta _{ij}$, 
but the output states are not.  To satisfy reversibility, this 
requires the {\it auxiliary} output states are orthogonal.

\[
\bk{Aux(i)}{Aux(j)}=\delta _{ij}
\]

Now, when we examine the TIDY condition, we find there does exist 
a unitary operator (and therefore a computable procedure) for 
tidying the computation, without losing the output. However, this 
tidying computation is not Bennett's procedure. If we were to 
clone the auxiliary output, and run the reverse operation, we 
would lose the output, and be left with the 'junk'!

It is possible\footnote{This section revises results reported in 
\cite{Mar01,Mar02}} to construct a modified version of Bennett's 
procedure for this case:
\begin{itemize}
\item Copy the joint states $\ket{O(i)}\ket{Aux(i)}$ to a second 
system.  As these states are mutually orthogonal, due to the 
orthogonality of the $\ket{Aux(i)}$ states, a controlled-NOT gate 
can be constructed working in this basis.
\item Conditionally reset the copy of the $\ket{Aux(i)}$ states 
using the originals:
\[
U_R:\ket{Aux(i)}\ket{Aux(i)}\rightarrow \ket{Aux(i)}\ket{Aux0}
\]
This is also possible, by constructing a controlled-NOT gate in 
the $\ket{Aux(i)}$ basis.
\item Reverse the computation, leaving only the copied output 
states.
\end{itemize}

However, this raises an interesting complication: do we know the 
two basis in question?  The strength of Bennett's procedure for 
classical computations is that we do not need to know the 
contents of the output or auxiliary bits.  For quantum 
computations we have no universal copying operation.  We can 
construct copying operations for any given orthogonal set of 
states, but this means that to implement the tidying procedure we 
need to know in advance at least the orthogonal basis in which 
the copying operation needs to take place.  Note, we do not need 
to know which specific output states $\ket{O(i)}\ket{Aux(i)}$ are 
associated with the specific input states $\ket{i}$, but we do 
need to know what the set of auxiliary output states 
$\ket{Aux(i)}$ will be and their correlation to the useful output 
states $\ket{O(i)}$.

It is not clear whether the operator can be constructed without 
explicitly computing each of the auxiliary output states on a 
second system - which may entail re-running the computation 
itself, for each input, and measuring the auxiliary output 
basis\footnote{A similar problem, of constructing a 
Controlled-NOT in a specific basis, exists for the case of 
orthogonal outputs, but in this case it seems reasonable assume 
that we do know in advance the set of the output states we are 
trying to compute, just not their relation to the input 
states.}.  This is precisely the problem that Bennett's procedure 
avoids for classical computation.

\subsection{Non-orthogonal Inputs}

The input states are a non-orthogonal set. This corresponds to 
Bob's position in the quantum distribution network of Figure 
\ref{fg:qnetwork}, and should be considered the most general case.

If we look at the requirements for a TIDY computation, this leads 
to:

\[
\bk{O(i)}{O(j)}=1
\]

The output is always the same, regardless of the input! Obviously 
for a computation to be meaningful, or non-trivial, at least some 
of the output states must depend in some way upon the particular 
input state. So in this case we can say there are NO procedures 
$F^{\prime \prime }$ that allow us to tidy our output from $F$. To 
state this exactly:
\begin{quotation}
There does not exist any non-trivial $\left( \ket{O(i)} \neq 
\ket{O(j)}\right) $ computations of the form

\[
G:\ket{i}\ket{Aux0}\ket{AuxO}
    \rightarrow \ket{i}\ket{Aux0}\ket{O(i)}
\]

for which $\bk{i}{j} \neq \delta_{ij}$.
\end{quotation}

It should be made clear: this does NOT mean useful quantum 
computations of the form

\[
F:\ket{i}\ket{Aux0}\rightarrow \ket{Aux(i)}\ket{O(i)}
\]

do not exist if $\bk{i}{j} \neq \delta_{ij}$ - simply that such 
computations cannot be TIDY. For such computations, not only are 
the bits used to store the auxiliary output unrecoverable, 
without dissipating energy, but also the input state cannot be 
recovered, except through losing the output. For our distributed 
network, this means that not only can Bob not tidy his 
computation, but he cannot restore Alice's data to the database.
\section{Conclusion}\label{conclusion}

For classical computations, Bennett provides a procedure by which 
a reversible computation may be made tidy. The great advantage of 
this procedure is that it does not require one to have advance 
knowledge of the output of the computation.  Instead, simply the 
knowledge of the successive logical operations comprising the 
computation is sufficient to implement the tidying procedure. 

However, this procedure rests upon the use of the Controlled-NOT 
operator, to clone the output states. For a general quantum 
computation this is not possible. Where the output states of a 
quantum computation form an orthogonal set, then a Controlled-NOT 
style gate can indeed be constructed. For such computations 
Bennett's procedure will work. If the output states of a quantum 
computation are not orthogonal, then the required generalisation 
of a Controlled-NOT gate would be a quantum state cloning device, 
and no such device exists. 

When the input states are orthogonal, then for non-orthogonal 
output states, the auxiliary outputs must be an orthogonal set. 
In this case we know that a unitary operator must exist for the 
tidying procedure. However, unlike for Bennett's procedure, it is 
unclear that there is a simple prescription for finding the 
tidying operation. If we are given the output auxiliary states, 
then we can derive a suitable operation. However, we will not, in 
general, know these states without having to actually perform the 
calculation with all the possible input states.  Thus this method 
defeats the purpose of trying to produce the tidy computation. 

Finally where input states are not orthogonal, then there is no 
possible tidying algorithm.  To recover the auxiliary bits 
requires a dissipative resetting.  This represents a fundamental 
energy requirement upon quantum computations that is not present 
in classical computations. 

\bibliographystyle{alpha}
\bibliography{paper5}

\begin{thebibliography}{LTV98}

\bibitem[Ben73]{Ben73}
C~H Bennett.
\newblock The logical reversibility of computation.
\newblock {\em IBM J Res Develop}, 17:525--532, 1973.

\bibitem[Lan61]{Lan61}
R~Landauer.
\newblock Irreversibility and heat generation in the computing process.
\newblock {\em IBM J Res Dev}, 5:183--191, 1961.
\newblock Reprinted in {\cite{LR90}}.

\bibitem[LR90]{LR90}
H~S Leff and A~F Rex, editors.
\newblock {\em Maxwell's Demon. Entropy, Information, Computing}. Adam Hilger,
  1990.
\newblock ISBN 0-7503-0057-4.

\bibitem[LTV98]{LTV98}
M~Li, J~Tromp, and P~Vitanyi.
\newblock Reversible simulation of irreversible computation by pebble games.
\newblock {\em Physica D}, 120(1-2):168--176, 1998.
\newblock quant-ph/9703009.

\bibitem[LV96]{LV96}
M~Li and P~Vitanyi.
\newblock Reversibility and adiabatic computation: Trading time and space for
  energy.
\newblock {\em Proc R Soc Lond A}, 452:769--789, 1996.

\bibitem[Mar01]{Mar01}
O~J~E Maroney.
\newblock Sameness and oppositeness in quantum information.
\newblock In {\em Proceedings 21st {ANPA} Conference}, 2001.
\newblock ISBN 0 9526215 6 8.

\bibitem[Mar02]{Mar02}
O~J~E Maroney.
\newblock {\em Information and Entropy in Quantum Theory}.
\newblock PhD thesis, Birkbeck College, University of London, 2002.
\newblock www.bbk.ac.uk/tpru/OwenMaroney/thesis/thesis.html.

\bibitem[WZ82]{WZ82}
W~K Wootters and W~H Zurek.
\newblock A single quantum cannot be cloned.
\newblock {\em Nature}, 299:802--803, 1982.

\end{thebibliography}

\end{document}